\documentclass[%
 reprint,
 amsmath,amssymb,
 prl,
]{revtex4-1}

\pdfoutput=1
\usepackage{graphicx}
\usepackage{dcolumn}
\usepackage{bm}
\usepackage{wrapfig}
\usepackage{amssymb}
\usepackage{epstopdf}

\begin{document}


\title{Exact, Dynamically Routable Current Propagation\\in Pulse-Gated Synfire Chains}

\author{Andrew T Sornborger}
\email{ats@math.ucdavis.edu}
\affiliation{Department of Mathematics, University of California, Davis, USA}%


\author{Louis Tao}
\email{taolt@mail.cbi.pku.cn.edu}
\affiliation{Center for Bioinformatics, National Laboratory of Protein Engineering and Plant Genetic Engineering, College of Life Sciences, and Center for Quantitative Biology, Peking University, Beijing, China}%

\date{\today}

\begin{abstract}
\noindent
Neural oscillations can enhance feature recognition \citep{AzouzGray2000}, modulate interactions between neurons \citep{WomelsdorfEtAl2007}, and improve learning and memory \citep{MarkowskaEtAl1995}. Simulational studies have shown that coherent oscillations give rise to windows in time during which information transfer can be enhanced in neuronal networks \citep{Abeles1982,LismanIdiart1995,SalinasSejnowski2001}. Unanswered questions are: 1) What is the transfer mechanism? And 2) how well can a transfer be executed? Here, we present a pulse-based mechanism by which graded current amplitudes may be {\it exactly} propagated from one neuronal population to another. The mechanism relies on the downstream gating of mean synaptic current amplitude from one population of neurons to another via a pulse. Because transfer is pulse-based, information may be dynamically routed through a neural circuit. We demonstrate the amplitude transfer mechanism in a realistic network of spiking neurons and show that it is robust to noise in the form of pulse timing inaccuracies, random synaptic strengths and finite size effects. In finding an exact, analytical solution to a fundamental problem of information coding in the brain, graded information transfer, we have isolated a basic mechanism that may be used as a building block for fast, complex information processing in neural circuits.
\end{abstract}

\pacs{Valid PACS appear here}
\maketitle

Understanding information coding is crucial to understanding how neural circuits and systems bind sensory signals into internal mental representations of the environment, process internal representations to make decisions, and translate decisions into motor activity. 

Classically, coding mechanisms have been shown to be related to neural firing rate \citep{AdrianZotterman1926}, population activity \citep{HubelWiesel1965,HubelWiesel1968,KaisslingPriesner1970}, and spike timing \citep{pmid8768391}. Firing rate \citep{AdrianZotterman1926} and population codes \citep{Knight1972,Knight2000,SirovichEtAl1999,Gerstner1995,BrunelHakim1999} are two different ways for a neural system to average spike number to represent graded stimulus information, with population codes capable of faster and more accurate processing since averaging is performed across many fast responding neurons. Thus population and temporal codes are capable of making use of the sometimes millisecond accuracy \citep{pmid8768391,pmid17805296,pmid23010933} of spike timing to represent signal dynamics.

The modern understanding of information coding in neural systems has become more nuanced, with investigations focusing on population oscillations \citep{QuianQuirogaPanzeri2013,pmid20725095}. Although classical mechanisms serve as their underpinnings, new mechanisms have been proposed for short-term memory \citep{LismanIdiart1995,JensenLisman2005,Goldman2008}, information transfer via spike coincidence \citep{Abeles1982,KonigEtAl1996,Fries2005} and information gating \citep{SalinasSejnowski2001,Fries2005,RubinTerman2004} that rely on gamma- and theta-band oscillations.

The Lisman-Idiart interleaved-memory (IM) model \citep{LismanIdiart1995}, Abeles's synfire network \citep{Abeles1982,KonigEtAl1996,pmid10591212,VogelsAbbott2005,pmid20365405,pmid21106815} and Fries's commu-nication--through--coherence (CTC) model \citep{Fries2005} all make use of the fact that well-timed oscillatory  excitations can provide windows in time during which spikes may be transferred between neurons. However, the extent to which information transfer can be enhanced by coherent oscillations has not been understood.

\begin{figure*}[]
  \includegraphics[width=\textwidth]{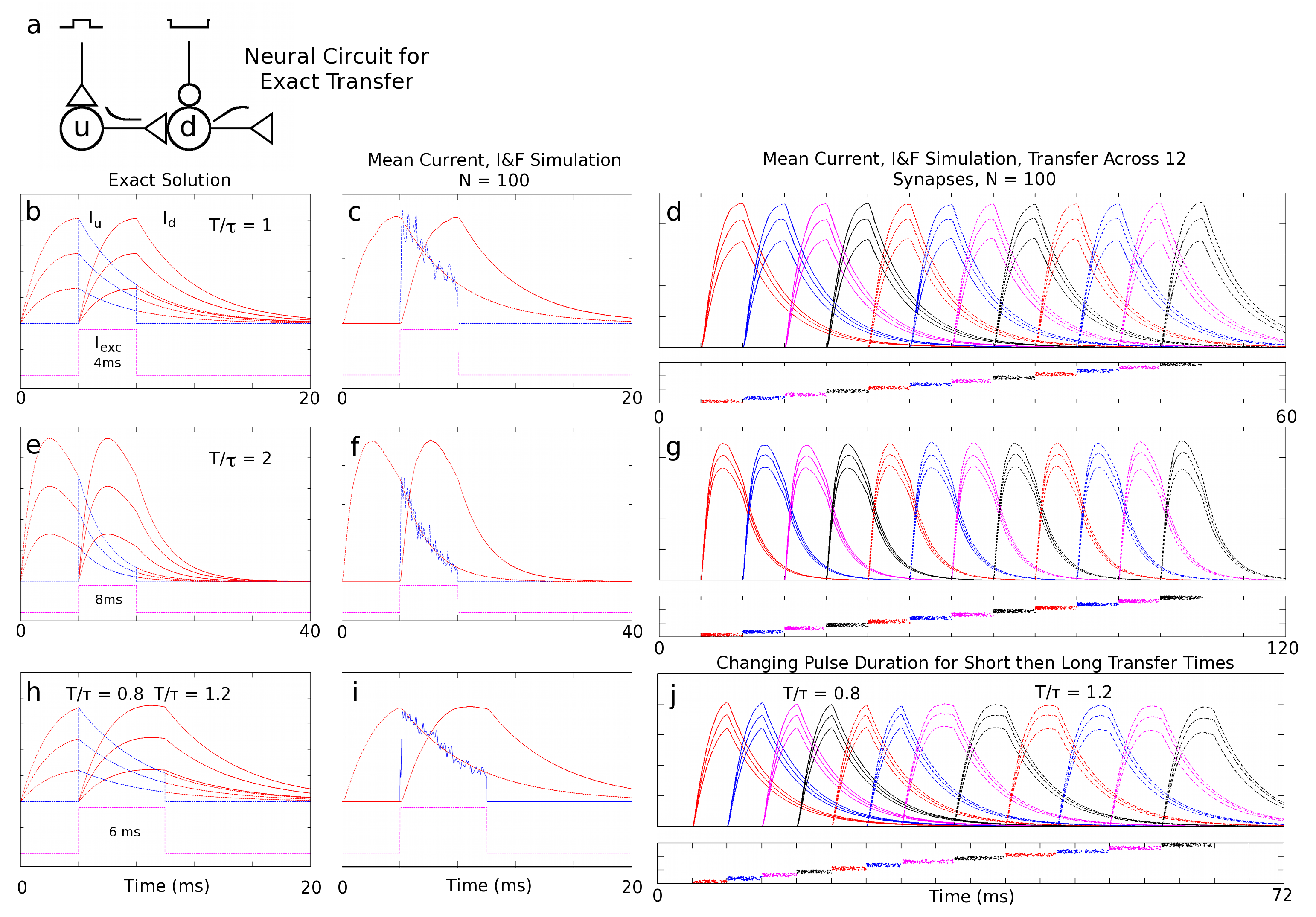}
  \caption{Exact current transfer - Mean-field and I\&F: a) Circuit diagram for the current transfer mechanism. Excitatory pulse gating the upstream ($u$) population and ongoing inhibition acting on the downstream ($d$) population. The upstream population excites the downstream population and transfers its current. b, e) Dynamics of a single current amplitude transfer with $T/\tau = 1$ and $T/\tau = 2$ with $\tau = 4$ ms. Dashed red traces represent the growth and exponential decay of three different current amplitudes, $I_u(t)$, in the upstream population. Dashed blue traces represent excitatory firing rates of the upstream population. Solid red traces represent the integration and subsequent decay of the downstream current, $I_d(t)$. Dashed magenta traces represent the excitatory gating pulse current from the upstream population. Magenta traces are displaced from zero for clarity. c, f) Dynamics of two $N = 100$ neuron populations of current-based, I\&F neurons showing current amplitude transfer averaged over 20 trials. d, g) Mean current amplitudes of twelve current-based, I\&F neuronal populations ($N = 100$ neurons/population, averaged over 20 trials) successively transferring their currents. Below each of these panels is a plot of one realization of the spike times for all neurons in each respective population. h) Dynamics of a single current amplitude transfer between two gating periods, with $T/\tau = 0.8$ and $T/\tau = 1.2$. i) Dynamics of two $N = 100$ neuron populations showing current amplitude transfer with the same gating periods as in i). j) Five graded current amplitude transfers between six populations with $T/\tau = 0.8$, followed by a transfer to $T/\tau = 1.2$, followed by five equally graded transfers with $T/\tau = 1.2$.}
\end{figure*}

Here, we show that information contained in the amplitude of a synaptic current may be {\it exactly} transferred from one neuronal population to another, as long as well-timed current pulses are injected into the populations. We derive this pulse-based transfer mechanism using mean-field equations for a current-based neural circuit (see circuit diagram in Fig. 1a). Graded current amplitudes are transferred between upstream and downstream populations: A gating pulse excites the upstream population into the firing regime generating a synaptic current in the downstream population. An ongoing inhibition keeps the downstream population silent until the feedforward synaptic current is integrated.

The downstream synaptic current is governed by
$$\tau \frac{d}{{dt}}{I_d} =  - {I_d} + S{m_u } \; ,$$ where $S$ is the synaptic coupling strength, $I_d$ is the downstream synaptic current, $m_u$ is the upstream firing rate and $\tau$ is a synaptic timescale.

During the current transfer epoch, $0 < t < T$, $$m_u = \left[ I_u(t) + I_0^{Exc} - I_0^{Inh} - g_0 \right]^+$$ (see Appendices for a complete discussion). The amplitude of the excitatory gating pulse, $I_0^{Exc}$ (Fig. 1b, dashed magenta), is set to $I_0^{Inh} + g_0$, allowing the downstream population to integrate the firing rate due to $I_u(t)$ exclusively (Fig. 1b, red). For an exponentially decaying upstream firing rate, $m_u = {I_u}\left( t \right) = A{e^{ - t/\tau }}$ (Fig. 1b, dashed blue), the integrated downstream current is ${I_d}\left( t \right) = SA\frac{t}{\tau }{e^{ - t/\tau }}$ (Fig. 1b, red).
During this time, even though the downstream current is integrated, the ongoing inhibition acts on the downstream population to keep it from spiking. 

For $T < t < 2T$, the downstream population is gated by an excitatory pulse, while the upstream population ceases firing. Thus, the downstream synaptic current decays exponentially from its value at $t = T$ (Fig. 1b, red). So that we have
$${I_d}\left( t \right) = SA\frac{T}{\tau }{e^{ - T/\tau }}{e^{ - \left( {t - T} \right)/\tau }}$$
and ${m_d} = {\left[ {{I_d}\left( t \right) + I_0^{Exc} - I_0^{Inh} - {g_0}} \right]^ + } = {I_d}\left( t \right)$. For exact transfer, we need ${I_d}\left( {t - T} \right) = {I_u}\left( t \right)$, requiring $S_{exact} = \frac{\tau }{T}{e^{ T/\tau }}$. 

This mechanism has a number of features that are represented in the analytic solution: 1) Exact transfer is possible for any $T$ and $\tau$. This means that transfer may be enacted on a wide range of time scales. This range is set roughly by the value of $S_{exact}$. Roughly, $0.1 < T/\tau < 4$ gives $S$ small enough that firing rates are not excessive. 2) $\tau$ sets the ``reoccupation time" of the upstream population. After one population has transferred its amplitude to another, the current amplitude must fall sufficiently close to zero for a subsequent exact transfer. Synapses mediated by AMPA (NMDA) allow repeated exact transfers in the gamma (theta) band, respectively. 3) Pulse-gating controls information flow, not information content. As an example, one upstream population may be synaptically connected to two (or more) downstream populations. The graded current amplitude may then be selectively transferred downstream depending on whether one, the other, or both downstream populations are pulsed appropriately. This allows the functional connectivity of neural circuits to be plastic and rapidly controllable by pulse generators. 4) $S_{exact}$ has an absolute minimum at $T/\tau = 1$, and, except at the minimum, there are always two values of $T/\tau$ that give the same value of $S$. This means, for instance, that an amplitude transferred via a short pulse may subsequently be transferred by a long pulse and vice versa (see Fig. 1h,i,j). Thus, not only may downstream information be multiplexed using pulse-based control, but the time scale of the mechanism may also be varied from transfer to transfer.

The means by which the mechanism can fail are also readily apparent: 1) The pulse might not be accurately timed. 2) Synaptic strengths might not be correct for exact transfer. 3) The amplitude of the excitatory pulse $I_0^{Exc}$ might not precisely cancel the effective threshold $I_0^{Inh} + g_0$. 4) The mean-field approximation might break down due to too few neurons in the neuronal populations.

In Fig. 1, we demonstrate the mechanism in both mean-field and spiking models. Fig. 1b and e show the exact, mean-field transfer solution for $T = \tau = 4$ ms and $T = 2\tau = 8$ ms. Fig. 1c and f show corresponding transfer between populations of $N = 100$ current-based, integrate-and-fire (I\&F) neurons. Fig. 1d and g show mean currents computed from simulations of I\&F networks with $N = 100$. Mean amplitude transfer for these populations is very nearly identical to the exact solution and, as may be seen, graded amplitudes are transferred across many synapses and are still very accurately preserved. Fig. 1h shows the exact mean-field transfer solution between populations gated for $T/\tau = 0.8$ and $T/\tau = 1.2$ with $\tau = 5$ ms. Fig. 1i shows the corresponding transfer between populations of $N = 100$ I\&F neurons. Fig. 1j shows how integration period, $T$, may be changed within a sequence of successive transfers.

\begin{figure}[ht]
  \includegraphics[width=\columnwidth]{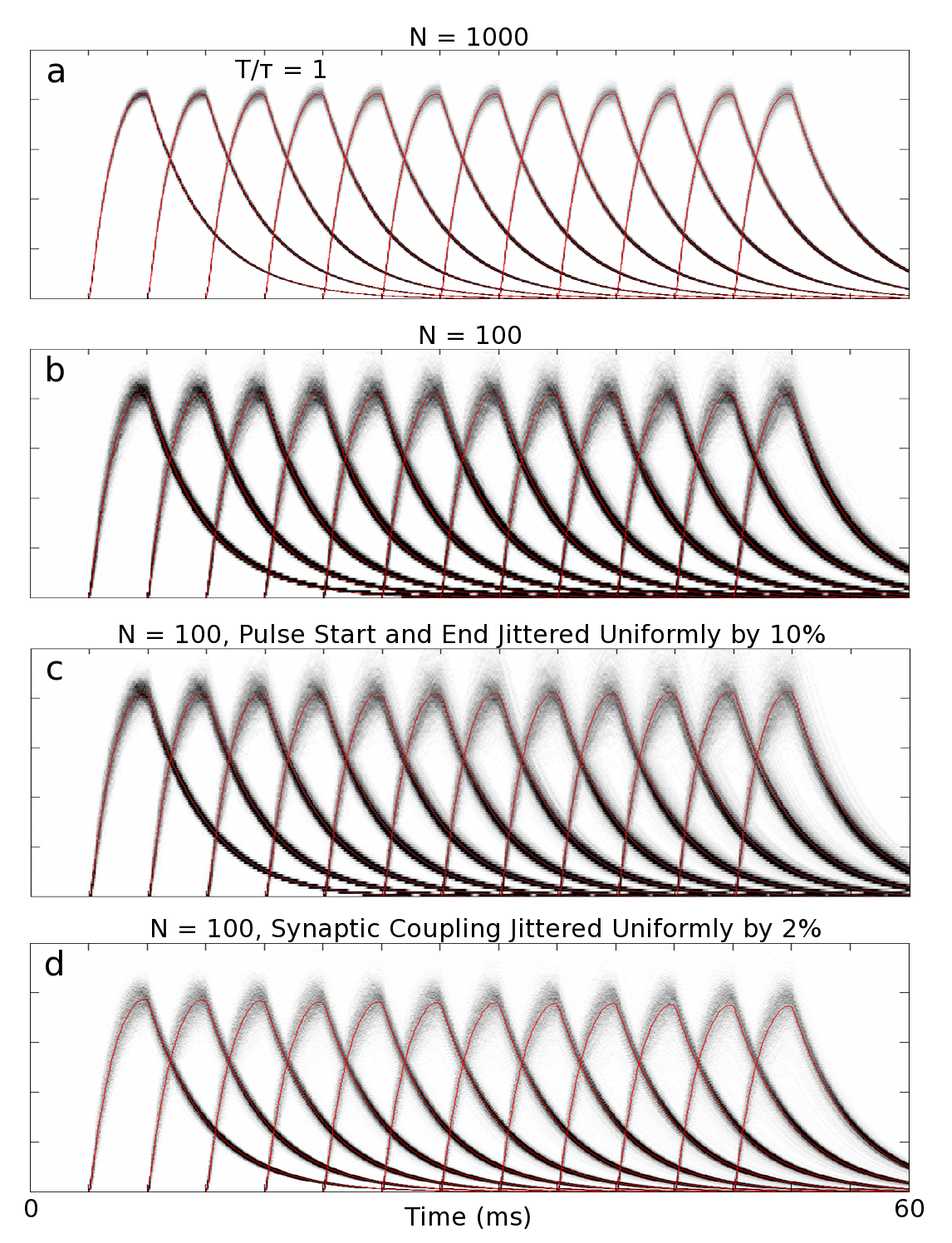}
    \caption{Sources of variability in the current transfer mechanism: Distributions of mean current amplitudes from current-based, I\&F neuronal simulations. a) $N = 1000$. b) $N = 100$. c) $N = 100$ with the start and end times of each pulse jittered uniformly by $10\%$ of the pulse width. d) $N=100$ with synaptic coupling jittered uniformly by 2\%. Gray scale: White denotes $0$ probability, Black denotes probability maximum. All distributions sum to unity along the vertical axis.}
\end{figure}

In Fig. 2, we investigate the mean current variability for the transfer mechanism in the spiking model due to the modes of failure discussed above for $T/\tau = 1$ with $\tau = 4$ ms. Fig. 2a shows the distribution of mean current amplitudes averaged over populations of $N = 1000$ neurons, calculated from $1000$ realizations. Fig. 2b shows the distribution with just $N = 100$. Clearly, more neurons per population gives less variability in the distribution. The signal-to-noise ratio (SNR) decreases as the square-root of the number of neurons per population, as would be expected. Thus, for circuits needing high accuracy, neuronal recruitment would increase the SNR. Fig. 2c shows the distribution for $N = 100$ with $10\%$ jitter in pulse start and end times. Fig. 2d shows the distribution for $N = 100$ with $2\%$ jitter in synaptic coupling.  Note that near $T/\tau = 1$, $S_{exact}$ varies slowly, thus the effect of both timing and synaptic coupling jitter on the stability of the transfer is minimal. Pulse timing, synaptic strengths, synaptic recruitment, and pulse amplitudes are regulated by neural systems. So mechanisms are already known that could allow networks to be optimized for graded current amplitude transfer.

The existence of exact graded transfer mechanisms, such as the one that we have found, goes a long way toward clarifying why the brain can be such a precise computer. It also points toward a natural modular organization wherein each circuit would be expected to have 1) sparsely coupled populations of neurons that encode information content, 2) pattern generators that provide accurately timed pulses to control information flow, and 3) regulatory mechanisms for maintaining optimal transfer.

A huge literature now exists implicating oscillations as an important mechanism for information coding. Our mechanism provides a fundamental building block with which graded information content may be encoded and transferred in current amplitudes, dynamically routed with coordinated pulses, and transformed and processed via synaptic weights. From this perspective, coherent oscillations may be an indication that a neural circuit is performing complex computations pulse by pulse. 

\begin{acknowledgments}
{\bf Acknowledgements}  L.T. thanks the UC Davis Mathematics Department for its hospitality. This work was supported by the Ministry of Science and Technology of China through the Basic Research Program (973) 2011CB809105 (L.T.) and the Natural Science Foundation of China grant 91232715 (L.T.). We thank Tim Lewis for reading and commenting on a draft of this paper.
\end{acknowledgments}

\appendix

\begin{center}
  {\bf Appendix}
\end{center}

\section{Circuit Model}
Our model circuit consists of a set of $j = 1, \dots, M$ populations, each with $i = 1, \dots, N$ neurons, with sparse feedforward connectivity. The probability that neuron $i$ in population $j$ synapses on neuron $k$ in population $j+1$ is $P_{ik} = p$. In our simulations, $pN = 80$.

The excitatory pulse on neurons in population $j$ is $$I_j^{Exc}(t) = (I_0^{Exc} + \epsilon) (\theta(t - jT) - \theta(t - (j+1)T)) \; ,$$ where $\theta(t)$ is the Heaviside step function: $\theta(t) = 0, \; t < 0$ and $\theta(t) = 1, \; t > 0$. The ongoing inhibitory current is $I_j^{Inh}(t) = I_0^{Inh}$. To avoid excessive synchronization, intrinsic noise is introduced in the excitatory pulse amplitude via $\epsilon$, where $\epsilon \sim N(0, \sigma^2)$, with $\sigma = g_{Leak}/50$.

Our network consists of a system of current-based, integrate-and-fire (I\&F) point neurons. Individual neurons have membrane potentials, $v_{i,j}$, described by
$$\frac{dv_{i,j}}{dt} = - g_{leak}(v_{i,j} - V_{Leak}) + I_{i,j}^{Total}$$ and
$$\tau\frac{dI_{i,j}^s}{dt} = -I_{i,j}^s + S \sum_i \sum_{k} \delta(t - t_{i,j-1}^k) \; ,$$
with $i = 1, \dots, N$ and $j = 1, \dots, M$, synaptic currents
$$I_{i,j}^{Total} = I_{i,j}^s + I_j^{Exc} - I_j^{Inh}$$ and $V_{Leak}$ is the leakage potential. Here, $\tau$ is a current relaxation timescale depending on the type of neuron (typical time constants are $\tau_{AMPA} \sim 3-11$ ms or $\tau_{NMDA} \sim 60-150$ ms). Individual spike times, $\{ t_{i,j}^k \}$, with $k$ denoting spike number, are determined by the time when the voltage $v_{i,j}$ reaches the threshold voltage $V_{Thres}$, at which time the voltage is reset to $V_{Reset}$. We use units in which only time retains dimension (in seconds): the leakage conductance is $g_{Leak} = 50/\mathrm{sec}$. We set $V_{Reset} = V_{Leak} = 0$ and normalize the membrane potential by the difference between the threshold and reset potentials, $V_{Thres} - V_{Reset} = 1$. For the simulations reported here, we use $I_0^{Exc} = 180$ and $I_0^{Inh} = 150$.

This network is effectively a synfire chain with prescribed pulse input \citep{Abeles1982,pmid10591212,VogelsAbbott2005,pmid20365405,KistlerGerstner2002}.

\section{Mean-field Equations}
Averaging (coarse-graining) spikes over time and over neurons in population $j$ (see, e.g. \citet{pmid12053156}) produces a mean firing rate equation given by
$${m_j} = \frac{{ - {g_{Total}}}}{{\ln \left( {\frac{{{{\left[ {{I_{j}^{Total}} - {g_{Total}}{V_{Thres}}} \right]}^ + }}}{{{g_{Total}}\left( {{V_{Thres}} - {V_{Reset}}} \right) + {{\left[ {{I_{j}^{Total}} - {g_{Total}}{V_{Thres}}} \right]}^ + }}}} \right)}} \; ,$$
where ${g_{Total}} = {g_{Leak}}$, $V_{Thres}$ is the threshold voltage, and
$${I_{j}^{Total}} = {I_j} + I_j^{Exc} - I_j^{Inh} \; .$$

The feedforward synaptic current, $I_{j+1}$, is described by
$$\tau \frac{d}{{dt}}{I_{j+1}} =  - {I_{j+1}} + S{m_j} \; .$$
The downstream population receives excitatory input, $m_j$, with synaptic coupling $S$ from the upstream population. 
As in the I\&F simulation, we set, $V_{Reset}=0$, and non-dimensionalize the voltage using $V_{Thres}-V_{Reset}=1$, so that 
$${m_j} = \frac{{ - {g_{Leak}}}}{{\ln \left( {\frac{{{{\left[ {{I_j^{Total}} - {g_{Leak}}} \right]}^ + }}}{{{g_{Leak}} + {{\left[ {{I_j^{Total}} - {g_{Leak}}} \right]}^ + }}}} \right)}} \; .$$
The f-I curve can be approximated by
\begin{eqnarray*}
m\left( {{I}} \right) & \approx & {\left[ {m\left( {{I_0}} \right) + m'\left( {{I_0}} \right)\left( {{I} - {I_0}} \right)} \right]^ + } \\
  & = & {\left[ {m'\left( {{I_0}} \right){I} - \left( {m'\left( {{I_0}} \right){I_0} - m\left( {{I_0}} \right)} \right)} \right]^ + }\\
  & \approx & {\left[ {{I} - {g_0}} \right]^ + } 
\end{eqnarray*}
near ${I} \approx {I_0}$, where  $m'\left( {{I_0}} \right) \approx 1$ (here the prime denotes differentiation), and letting ${g_0} = m'\left( {{I_0}} \right){I_0} - m\left( {{I_0}} \right)$ be the effective threshold in the linearized f-I curve.

\section{Exact Transfer}
We consider transfer between an upstream population and a downstream population, denoted by $j = u$ and $j+1 = d$.

For the downstream population, for $t < 0$, $I_d = 0$. This may be arranged as an initial condition or by picking a sufficiently large $I_0^{Inh}$, with 
$$\tau \frac{d}{{dt}}{I_d} =  - {I_d} + S{\left[ {{I_u}\left( t \right) - I_0^{Inh} - {g_0}} \right]^+ } \; .$$
At $t = 0$, the excitatory gating pulse is turned on for the upstream population for a period $T$, so that for $0 < t < T$, the synaptic current of the downstream population obeys
$$\tau \frac{d}{{dt}}{I_d} =  - {I_d} + S{\left[ {{I_u}\left( t \right) + I_0^{Exc} - I_0^{Inh} - {g_0}} \right]^+ } \; .$$
Therefore, we set the amplitude of the excitatory gating pulse to be $I_0^{Exc}=I_0^{Inh} + g_0$ to cancel the threshold. Making the ansatz ${I_u}\left( t \right) = A{e^{ - t/\tau }}$, we integrate $$\tau \frac{d}{{dt}}{I_d} =  - {I_d} + S{I_u}$$ to obtain the expression
$${I_d}\left( t \right) = SA\frac{t}{\tau }{e^{ - t/\tau }}, \; 0 < t < T \; .$$
During this time, ongoing inhibition is acting on the downstream population to keep it from spiking, i.e., we have
$${m_d}\left( t \right) = {\left[ {{I_d}\left( t \right) - I_0^{Inh} - {g_0}} \right]^ + } = 0 \; .$$
For $T < t < 2T$, the downstream population is gated by an excitatory pulse, while the upstream population is silenced by ongoing inhibition. The downstream synaptic current obeys
$$\tau \frac{d}{{dt}}{I_d} =  - {I_d}$$ with $${I_d}\left( T \right) = SA\frac{T}{\tau }{e^{ - T/\tau }}, \; T < t < 2T \; .$$
so that we have
$${I_d}\left( t \right) = SA\frac{T}{\tau }{e^{ - T/\tau }}{e^{ - \left( {t - T} \right)/\tau }}, \; T < t < 2T$$
and $${m_d} = {\left[ {{I_d}\left( t \right) + I_0^{Exc} - {g_0}} \right]^ + } = {I_d}\left( t \right) \; .$$ For exact transfer, we need ${I_d}\left( {t - T} \right) = {I_u}\left( t \right)$, therefore we write 
$$SA\frac{T}{\tau }{e^{ -{T}/\tau }} = A \; .$$
So we have exact transfer with $S_{exact} = \frac{\tau }{T}{e^{ T/\tau }}$. 

To recap, we have the solution
$${I_d}\left( t \right) = \left\{ {\begin{array}{*{20}{ll}}
   {SA\frac{t}{\tau }{{\mathop{\rm e}\nolimits} ^{ - t/\tau }}}, & 0 < t < T  \\
   {SA\frac{T}{\tau }{{\mathop{\rm e}\nolimits} ^{ - t/\tau }}}, & T < t < \infty
\end{array}} \right.$$
and
$${m_d}\left( t \right) = \left\{ {\begin{array}{*{20}{ll}}
   0, & 0 < t < T  \\
  SA\frac{T}{\tau }{{\mathop{\rm e}\nolimits}^{ - t/\tau }}, & T < t < 2T \\ 
   0, & 2T < t < \infty
\end{array}} \right. \; .$$

\bibliographystyle{unsrtnat}
\bibliography{Biblio}

\begin{thebibliography}{31}
\providecommand{\natexlab}[1]{#1}
\providecommand{\url}[1]{\texttt{#1}}
\expandafter\ifx\csname urlstyle\endcsname\relax
  \providecommand{\doi}[1]{doi: #1}\else
  \providecommand{\doi}{doi: \begingroup \urlstyle{rm}\Url}\fi

\bibitem[Azouz and Gray(2000)]{AzouzGray2000}
R.~Azouz and C.M. Gray.
\newblock Dynamic spike threshold reveals a mechanism for synaptic coincidence
  detection in cortical neurons in vivo.
\newblock \emph{Proc. Natl. Acad. Sci. USA}, 97:\penalty0 8110--8115, 2000.

\bibitem[Womelsdorf et~al.(2007)Womelsdorf, Schoffelen, Oostenveld, Singer,
  Desimone, Engel, and Fries]{WomelsdorfEtAl2007}
T.~Womelsdorf, J.M. Schoffelen, R.~Oostenveld, W.~Singer, R.~Desimone, A.K.
  Engel, and P.~Fries.
\newblock Modulation of neuronal interactions through neuronal synchronization.
\newblock \emph{Science}, 316:\penalty0 1609--1612, 2007.

\bibitem[Markowska et~al.(1995)Markowska, Olton, and Givens]{MarkowskaEtAl1995}
A.L. Markowska, D.S. Olton, and B.~Givens.
\newblock {Cholinergic manipulations in the medial septal area: Age-related
  effects on working memory and hippocampal electrophysiology}.
\newblock \emph{J. Neurosci.}, 15:\penalty0 2063--2073, 1995.

\bibitem[Abeles(1982)]{Abeles1982}
M.~Abeles.
\newblock Role of the cortical neuron: Integrator or coincidence detector?
\newblock \emph{Isr. J. Med. Sci.}, 18:\penalty0 83--92, 1982.

\bibitem[Lisman and Idiart(1995)]{LismanIdiart1995}
J.E. Lisman and M.A. Idiart.
\newblock Storage of $7\pm2$ short-term memories in oscillatory subcycles.
\newblock \emph{Science}, 267:\penalty0 1512--1515, 1995.

\bibitem[Salinas and Sejnowski(2001)]{SalinasSejnowski2001}
E.~Salinas and T.J. Sejnowski.
\newblock Correlated neuronal activity and the flow of neural information.
\newblock \emph{Nat. Rev. Neurosci.}, 2:\penalty0 539--550, 2001.

\bibitem[Adrian and Zotterman(1926)]{AdrianZotterman1926}
E.D. Adrian and Y.~Zotterman.
\newblock The impulses produced by sensory nerve-endings: Part {II}. {T}he
  response of a single-end organ.
\newblock \emph{J. Physiol.}, 61:\penalty0 151--171, 1926.

\bibitem[Hubel and Wiesel(1965)]{HubelWiesel1965}
D.H. Hubel and T.N. Wiesel.
\newblock Receptive fields and functional architecture in two non striate
  visual areas (18 and 19) of the cat.
\newblock \emph{J. Neurophysiol.}, 28:\penalty0 229--289, 1965.

\bibitem[Hubel and Wiesel(1968)]{HubelWiesel1968}
D.H. Hubel and T.N. Wiesel.
\newblock Receptive fields and functional architecture of monkey striate
  cortex.
\newblock \emph{J. Physiol.}, 195:\penalty0 215--243, 1968.

\bibitem[Kaissling and Priesner(1970)]{KaisslingPriesner1970}
K.E. Kaissling and E.~Priesner.
\newblock Smell threshold of the silkworm.
\newblock \emph{Naturwissenschaften}, 57:\penalty0 23--28, 1970.

\bibitem[Bair and Koch(1996)]{pmid8768391}
W.~Bair and C.~Koch.
\newblock {{T}emporal precision of spike trains in extrastriate cortex of the
  behaving macaque monkey}.
\newblock \emph{Neural Comput}, 8:\penalty0 1185--1202, 1996.

\bibitem[Knight(1972)]{Knight1972}
B.W. Knight.
\newblock Dynamics of encoding in a population of neurons.
\newblock \emph{J. Gen. Physiol.}, 59:\penalty0 734--766, 1972.

\bibitem[Knight(2000)]{Knight2000}
B.W. Knight.
\newblock Dynamics of encoding in a population of neurons: Some general
  mathematical features.
\newblock \emph{Neural Comput.}, 12:\penalty0 473--518, 2000.

\bibitem[Sirovich et~al.(1999)Sirovich, Knight, and Omurtag]{SirovichEtAl1999}
L.~Sirovich, B.W. Knight, and A.~Omurtag.
\newblock Dynamics of neuronal populations: The equilibrium solution.
\newblock \emph{SIAM J. Appl. Math.}, 60:\penalty0 2009--2028, 1999.

\bibitem[Gerstner(1995)]{Gerstner1995}
W.~Gerstner.
\newblock Time structure of the activity in neural network models.
\newblock \emph{Phys. Rev. E}, 51:\penalty0 738--758, 1995.

\bibitem[Brunel and Hakim(1999)]{BrunelHakim1999}
N.~Brunel and V.~Hakim.
\newblock Fast global oscillations in networks of integrate-and-fire neurons
  with low firing rates.
\newblock \emph{Neural Comput.}, 11:\penalty0 1621--1671, 1999.

\bibitem[Butts et~al.(2007)Butts, Weng, Jin, Yeh, Lesica, Alonso, and
  Stanley]{pmid17805296}
D.~A. Butts, C.~Weng, J.~Jin, C.~I. Yeh, N.~A. Lesica, J.~M. Alonso, and G.~B.
  Stanley.
\newblock {{T}emporal precision in the neural code and the timescales of
  natural vision}.
\newblock \emph{Nature}, 449:\penalty0 92--95, 2007.

\bibitem[Varga et~al.(2012)Varga, Golshani, and Soltesz]{pmid23010933}
C.~Varga, P.~Golshani, and I.~Soltesz.
\newblock {{F}requency-invariant temporal ordering of interneuronal discharges
  during hippocampal oscillations in awake mice}.
\newblock \emph{Proc. Natl. Acad. Sci. U.S.A.}, 109:\penalty0 E2726--2734,
  2012.

\bibitem[Quian~Quiroga and Panzeri(2013)]{QuianQuirogaPanzeri2013}
R.~Quian~Quiroga and S.~Panzeri, editors.
\newblock \emph{Principles of neural coding}.
\newblock CRC Press, London, 2013.

\bibitem[Kumar et~al.(2010)Kumar, Rotter, and Aertsen]{pmid20725095}
A.~Kumar, S.~Rotter, and A.~Aertsen.
\newblock {{S}piking activity propagation in neuronal networks: {R}econciling
  different perspectives on neural coding}.
\newblock \emph{Nat. Rev. Neurosci.}, 11:\penalty0 615--627, 2010.

\bibitem[Jensey and Lisman(2005)]{JensenLisman2005}
O.~Jensey and J.E. Lisman.
\newblock Hippocampal sequence-encoding driven by a cortical multi-item working
  memory buffer.
\newblock \emph{Trends Neurosci.}, 28:\penalty0 67--72, 2005.

\bibitem[Goldman(2008)]{Goldman2008}
M.S. Goldman.
\newblock {Memory without feedback in a neural network}.
\newblock \emph{Neuron}, 61:\penalty0 621--634, 2008.

\bibitem[K\"onig et~al.(1996)K\"onig, Engel, and Singer]{KonigEtAl1996}
P.~K\"onig, A.K. Engel, and W.~Singer.
\newblock Integrator or coincidence detector? {T}he role of the cortical neuron
  revisited.
\newblock \emph{Trends Neurosci.}, 19:\penalty0 130--137, 1996.

\bibitem[Fries(2005)]{Fries2005}
P.~Fries.
\newblock A mechanism for cognitive dynamics: Neuronal communication through
  neuronal coherence.
\newblock \emph{Trends Cogn. Sci.}, 9:\penalty0 474--480, 2005.

\bibitem[Rubin and Terman(2004)]{RubinTerman2004}
J.E. Rubin and D.~Terman.
\newblock High frequency stimulation of the subthalamic nucleus eliminates
  pathological thalamic rhythmicity in a computational model.
\newblock \emph{J. Comp. Neurosci.}, 16:\penalty0 211--235, 2004.

\bibitem[Diesmann et~al.(1999)Diesmann, Gewaltig, and Aertsen]{pmid10591212}
M.~Diesmann, M.~O. Gewaltig, and A.~Aertsen.
\newblock {{S}table propagation of synchronous spiking in cortical neural
  networks}.
\newblock \emph{Nature}, 402:\penalty0 529--533, 1999.

\bibitem[Vogels and Abbott(2005)]{VogelsAbbott2005}
T.~P. Vogels and L.~F. Abbott.
\newblock {{S}ignal propagation and logic gating in networks of
  integrate-and-fire neurons}.
\newblock \emph{J Neurosci}, 25:\penalty0 10786--10795, 2005.

\bibitem[Shinozaki et~al.(2010)Shinozaki, Okada, Reyes, and
  Cateau]{pmid20365405}
T.~Shinozaki, M.~Okada, A.~D. Reyes, and H.~Cateau.
\newblock {{F}lexible traffic control of the synfire-mode transmission by
  inhibitory modulation: nonlinear noise reduction}.
\newblock \emph{Phys Rev E Stat Nonlin Soft Matter Phys}, 81:\penalty0 011913,
  2010.

\bibitem[Kremkow et~al.(2010)Kremkow, Aertsen, and Kumar]{pmid21106815}
J.~Kremkow, A.~Aertsen, and A.~Kumar.
\newblock {{G}ating of signal propagation in spiking neural networks by
  balanced and correlated excitation and inhibition}.
\newblock \emph{J. Neurosci.}, 30:\penalty0 15760--15768, 2010.

\bibitem[Kistler and Gerstner(2002)]{KistlerGerstner2002}
W.M. Kistler and W.~Gerstner.
\newblock {Stable propagation of activity pulses in populations of spiking
  neurons}.
\newblock \emph{Neural Computation}, 14:\penalty0 987--997, 2002.

\bibitem[Shelley and McLaughlin(2002)]{pmid12053156}
M.~Shelley and D.~McLaughlin.
\newblock {{C}oarse-grained reduction and analysis of a network model of
  cortical response: {I}. {D}rifting grating stimuli}.
\newblock \emph{J Comput Neurosci}, 12\penalty0 (2):\penalty0 97--122, 2002.

\end{thebibliography}

\end{document}